\documentclass[]{spie}  

\usepackage{amsmath,amsfonts,amssymb}
\usepackage{graphicx}
\usepackage{caption}
\usepackage{subcaption}
\usepackage{multirow}
\usepackage[colorlinks=true, allcolors=blue]{hyperref}

\title{Glow reduction of ultra-low noise LmAPDs: towards photon counting infrared arrays}

\author[a]{Guillaume Huber}
\author[a]{Michael Bottom}
\author[a,b]{Charles-Antoine Claveau}
\author[a]{Shane Jacobson}
\author[a]{Matthew Newland}
\author[c]{Ian Baker}
\author[c]{Keith Barnes}
\author[c]{Matthew Hicks}
\affil[a]{Institute for Astronomy, University of Hawai’i, Hilo, HI 96720, USA}
\affil[b]{University of California, Berkeley, CA 94720, USA}
\affil[c]{Leonardo UK Ltd., Southampton, S015 0LG, UK}

\authorinfo{Further author information: (Send correspondence to Guillaume Huber)\\Guillaume Huber: E-mail: ghuber@hawaii.edu}

\begin{document} 
\maketitle

\begin{abstract}
Spectroscopy and direct-imaging of ultra-faint targets such as Earth-like exoplanets and high redshift galaxies are among the primary goals of upcoming large scale astronomy projects like the Habitable World Observatory (HWO). Such objectives pose extreme instrumental challenges, in particular on detectors where dark currents lower than 1 e-/pixel/kilosecond and read noise less than 1 e-/pixel/frame will have to be achieved on large format arrays. Some technologies meet these requirements at optical wavelengths, but none do in the infrared.\\
With this goal in mind, the University of Hawaii has partnered with Leonardo to develop linear-mode avalanche photodiodes (LmAPDs). In this paper, we report recent tests performed on LmAPDs, where we measure a ROIC glow of $\sim 0.01$ e-/pixel/frame, without which the intrinsic dark current is essentally zero ($<0.1$ e-/pixel/kilosecond). We show that at high gain, these devices are capable of detecting single photons. 
\end{abstract}


\section{INTRODUCTION}
\label{sec:intro}  

The James Webb Space Telescope has demonstrated the capability to characterize a diversity of exoplanets and detect new chemical signatures through transmission spectroscopy \cite{2023Natur.614..659R, 2023Natur.614..649J}. However, in order to detect bio-signatures on Earth-like exoplanets, more sensitive direct-imaging instruments must be developed. The Astro2020 Decadal survey \cite{2021pdaa.book.....N} strongly advocates for a coronographic $\sim$6-meter infrared (IR)/optical/ultraviolet space telescope with this specific science goal: the Habitable World Observatory. Such a telescope will have the challenge of measuring signals of only a few photons per pixel per hour, which unsurprisingly sets extreme requirements on detectors.

Significant sensor maturation for the Roman telescope has resulted in EMCCDs that can meet the requirements of such missions \cite{2016JATIS...2a1007H}. However, in the near-infrared (NIR), where many interesting molecular bands (H2O, CO2, CH4, etc) of biosignature gases are located, there is no flight-qualified detector technology that meets these needs. The state-of-the-art HAWAII-xRG arrays operate in the NIR and meet the requirement in terms of dark current \cite{2020JATIS...6a6001R}. However, the readout noise of these detectors is up to two orders of magnitude too high \cite{10.1117/12.787971} and there is no imminent technological path to overcoming this.

Over the last decade, with NASA support, UK based Leonardo and the IfA have been maturing an alternate technology for NIR astronomy; linear mode avalanche photodiode arrays (LmAPDs) (\textit{e.g.} the SAPHIRA detectors \cite{2016SPIE.9915E..0NA, 2018JATIS...4b6001G}) which have shown very promising performances. The latest generation of these sensors offers the potential for noise-free signal amplification before the readout noise penalty. This effectively reduces the read noise proportionally to the signal amplification factor, a.k.a the gain. These avalanche photodiodes are constructed out of HgCdTe, and so inherit the excellent infrared material properties (e.g., high QE and tunable wavelength cutoff) with conventional cryogenic operation \cite{2017SPIE10177E..26B}.

Preliminary tests and characterizing of the LmAPDs arrays have been conducted over the last years on prototype detectors. The latest results were presented in Reference\cite{2022SPIE12191E..0ZC} (hereafter C22), where notably a dark current of 3e-/pixel/kilosecond is reported. C22 attributes this value to a ROIC glow of 0.08e-/pixel/frame, which is suspected to virtually be the main noise contributor and is therefore essential to measure. In this paper, we characterize a new chip of which the pixel structure has been modified to reduce the glow. We measure a glow level of 0.012 e-/pixel/frame and an intrinsic dark current below 0.1 e-/pixel/kilosecond. Finally, we report individual photon detection in the lab.

The structure of this paper is the following: in Section \ref{lmapd} we summarily describe the LmAPD technology and in particular its architecture at the pixel level to relate it to glow and dark current. The measurements of these effects are presented and analyzed in Section \ref{glow} after introducing our experimental test bench. Once the performance of the detector is established, we operate in an ultra-low flux regime to study its ability to detect individual photons, which is shown in Section \ref{single}. Finally, Section \ref{sec:summary} provides a summary of our results and discusses the next steps planned in the development and characterization of these detectors.

\section{LmAPD detector technology and testing}\label{lmapd}

We here briefly introduce mercury cadmium telluride (HgCdTe) as a semiconductor for NIR detector technology, and how LmAPDs make use of it in their design. We then go over the architecture of the LmAPDs at the pixel level. Readers looking for a more detailed presentation of the LmAPD arrays may wish to read Ref.\cite{2022SPIE12183E..17Z}

\subsection{HgCdTe and the avalanche process}

\begin{figure*}
\begin{center}
\begin{tabular}{c}
\includegraphics[height=6cm]{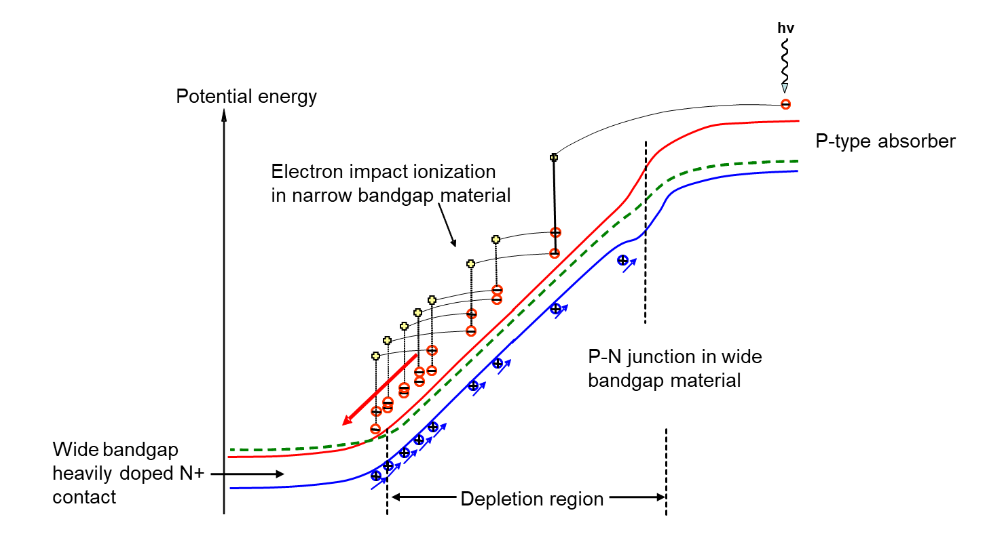}
\end{tabular}
\end{center}
\caption 
{ \label{fig:avalanche}
Potential energy diagram illustrating the history-dependent avalanche process.}
\end{figure*}

HgCdTe is widespread in the detector industry due to its high quantum efficiency in the optical-NIR and also its large, tunable spectral range \cite{2011OptLT..43.1358S}. At low temperature, its electron mobility can be very high compared to most semiconductors, making it a suitable material for avalanche photodiodes (APDs). The APD amplification process is almost noise-free with HgCdTe as the electrons do not experience phonon interactions or scattering during their multiplication. Additionally, the amplification happens before the noise penalty, essentially decreasing the effective read noise relative to the signal. Fig.\ref{fig:avalanche} illustrates the avalanche mechanism of LmAPDs.

\subsection{LmAPD architecture}

LmAPDs are grown via metal organic vapour phase epitaxy (MOVPE) at Leonardo. This manufacturing process uses a precise control of the semiconductors thickness with variable profiles enabling complex diode structures. 

\begin{figure}
\begin{center}
\begin{tabular}{c}
\includegraphics[height=5.4cm]{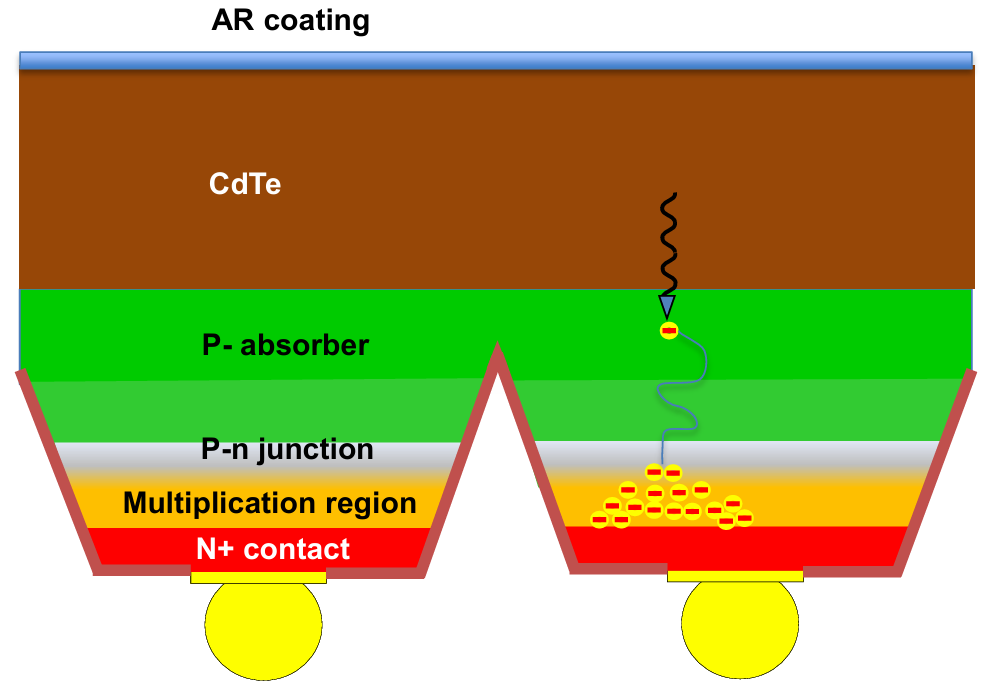}
\end{tabular}
\end{center}
\caption 
{ \label{fig:diode}
Schematic of the diode structure illustrating the conversion of a photon into a photoelectron and then its multiplication in the amplification process.}
\end{figure}

The architecture of the LmAPD at the pixel level is shown on Fig.\ref{fig:diode}. The top CdTe layer cuts off wavelengths below $0.8 \mu m$. Incident photons with $\lambda<2.5\mu m$ get converted into electrons in the P-type absorber layer. The generated photoelectrons diffuse through the p-n junction and then experience the avalanche process induced by the electric field of the multiplication region. At the bottom of the diode is a contact pad and an indium bump to connect it with the readout integrated circuits (ROIC) which collects the avalanche electrons and converts them into a electric signal.

The conic structure of the individual diodes allows for electrical isolation between pixels which reduces the inter-pixel cross-talk. The absorber is continuous, resulting in a complete photon absorption within the pixel.

\subsection{Experimental setup}

\begin{figure*}[!tbp]
\centering
    \begin{subfigure}[b]{0.45\textwidth}
    \centering
    \includegraphics[width=\textwidth]{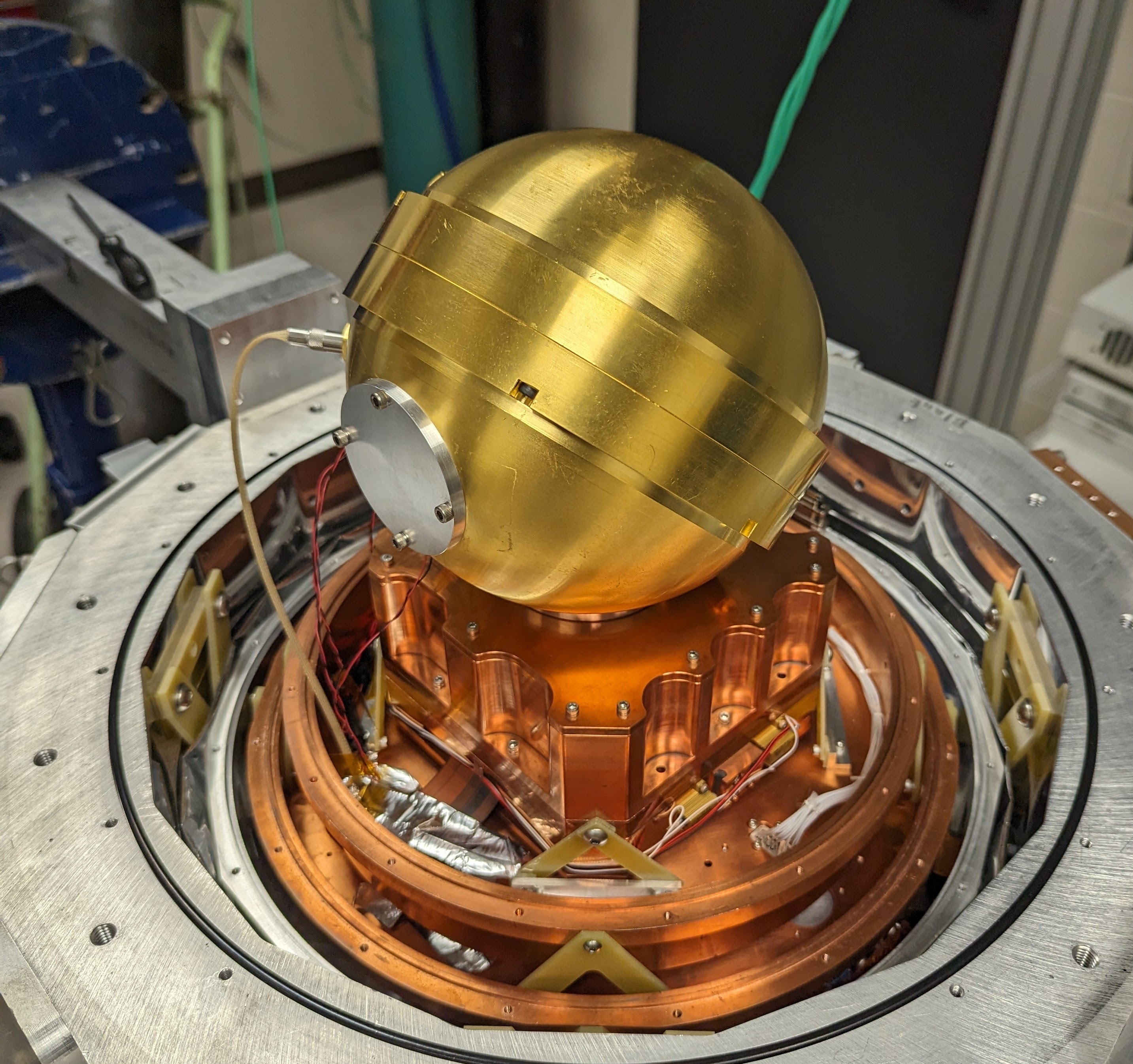}
    \label{subfig:sphere}
  \end{subfigure}
  \hfill
  \begin{subfigure}[b]{0.45\textwidth}
    \centering
    \includegraphics[width=\textwidth]{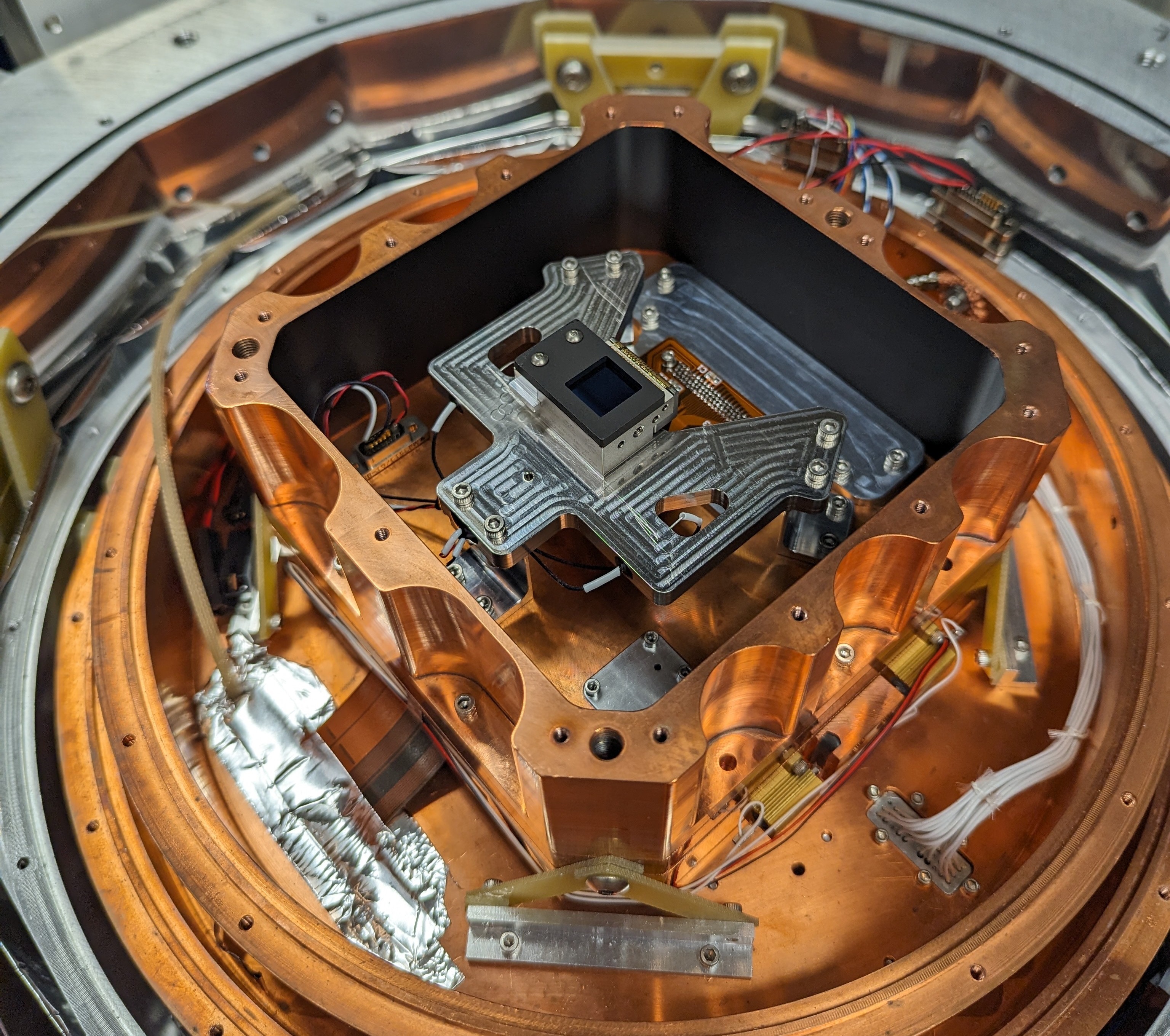}
    \label{fig:mask}
  \end{subfigure}
  \caption{Pictures of the detector cryogenic vacuum chamber. \textit{Left:} The fiber fed cryogenic integrating sphere mounted on the detector; \textit{right:} the inner detector housing and mounting structure. The matte mask is fixed above the detector.}
  \label{fig:setup}
\end{figure*}

Our laboratory setup is composed of a cryogenic chamber which can be cooled down to 40K with mK stability and pumped down to $<10^{-6}$ Torr. We use a broadband light source which we can control the flux via a variable optical attenuator. The light then enters the vacuum chamber and goes through a cryogenic integrating sphere (\textit{cf.} Fig.\ref{fig:setup}), providing a flat field onto the detector with minimal thermal emission. In order to monitor any parasitic signal (\textit{eg.} thermal emission within the chamber), a dark mask is placed very near the detector, under which $\sim$20\% of the pixels are blocked from any external light as seen on Fig.\ref{fig:setup}.

After the ROIC, the readout chain is composed of a SIDECAR ASIC \cite{2003SPIE.4841..782L} which is a single-chip microcontroller that can operate the detector and digitize its data. It interfaces to our control computer via a MACIE (Multi-purpose ASIC Control and Interface Electronics) controller card. Finally, two external power supplies are used to power the electronics and bias the detector. More information about the readout chain and data aquisition process can be found in C22.

The performance of the detector varies with several properties, in particular with its temperature and with the bias voltage, which sets the amplification gain. In this paper, the tests are conducted at 50K, with a bias of 4V. The conversion gain has been measured in these conditions via photon transfer curve (value in Tab.\ref{tab:summary}) so that results can be expressed in units of photoelectrons. It is specified when the detector is operated under different conditions.

\section{Glow and dark-current tests}\label{glow}

\begin{figure*}[!tbp]
\centering
    \begin{subfigure}[b]{0.4\textwidth}
    \centering
    \includegraphics[width=\textwidth]{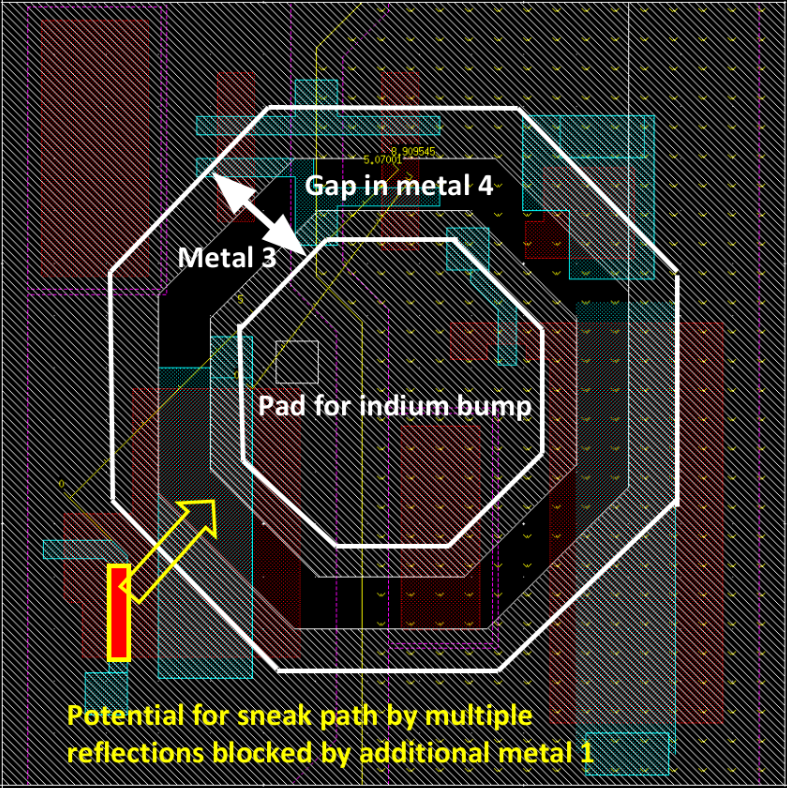}
  \end{subfigure}
  \hfill
  \begin{subfigure}[b]{0.44\textwidth}
    \centering
    \includegraphics[width=\textwidth]{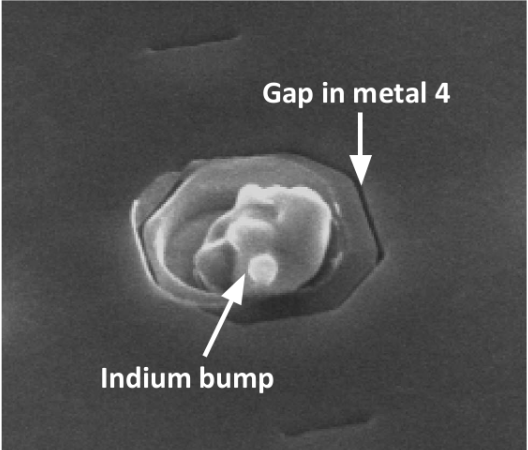}
  \end{subfigure}
  \caption{Potential path for glow propagation within the pixel. \textit{Left:} Schematic view from above the pixel at the node level. The red component is a JFET from the source follower, potentially glowing through the metal gap onto the higher photo-sensitive region; \textit{Right:} Close-up picture where the metal gap can be seen.}
  \label{fig:glow path}
\end{figure*}

The long-wavelength sensitivity of LmAPDs makes them vulnerable to self-generated emissions. While the intrinsic dark current depends on the exposure time, the glow contribution is a function of the number of frames taken and has been reported in several infrared sensors \cite{2020JATIS...6a6001R, 2020SPIE11454E..31L}. C22 concluded that the ROIC glow was potentially responsible for a 0.08 e-/pixel/frame signal revealing itself when a read frame is taken, making it the major source of noise of LmAPDs. The physical origin of the glow was suspected to be one of the JFETs of the source follower leaking through a gap in the metal architecture at the pixel node level as illustrated in Fig.\ref{fig:glow path}. This has been corrected for in the new chip by adding an extra layer of material to block the leakage path.

\begin{figure*}[!tbp]
\centering
    \begin{subfigure}[b]{0.48\textwidth}
    \centering
    \includegraphics[width=\textwidth]{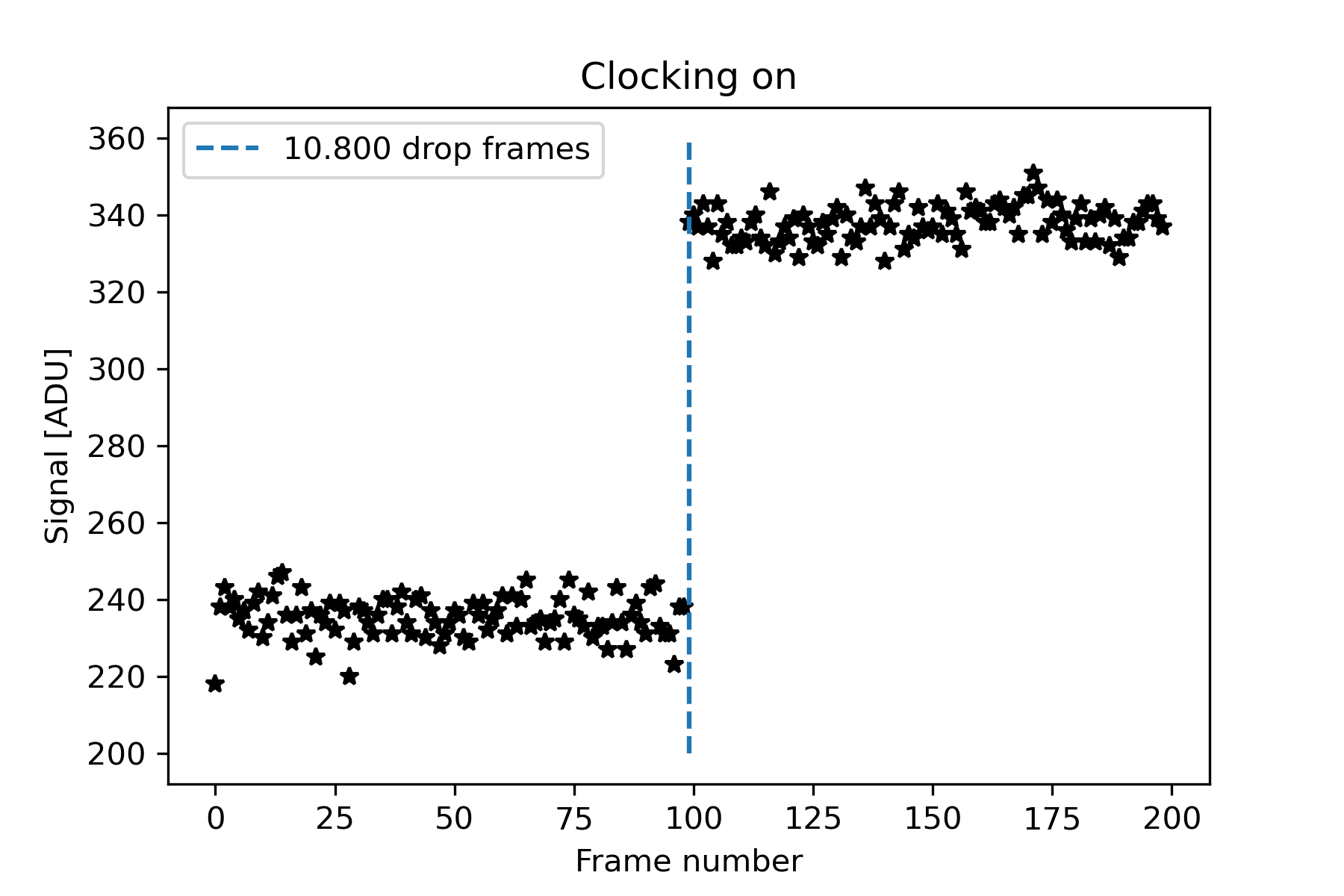}
  \end{subfigure}
  \hfill
  \begin{subfigure}[b]{0.48\textwidth}
    \centering
    \includegraphics[width=\textwidth]{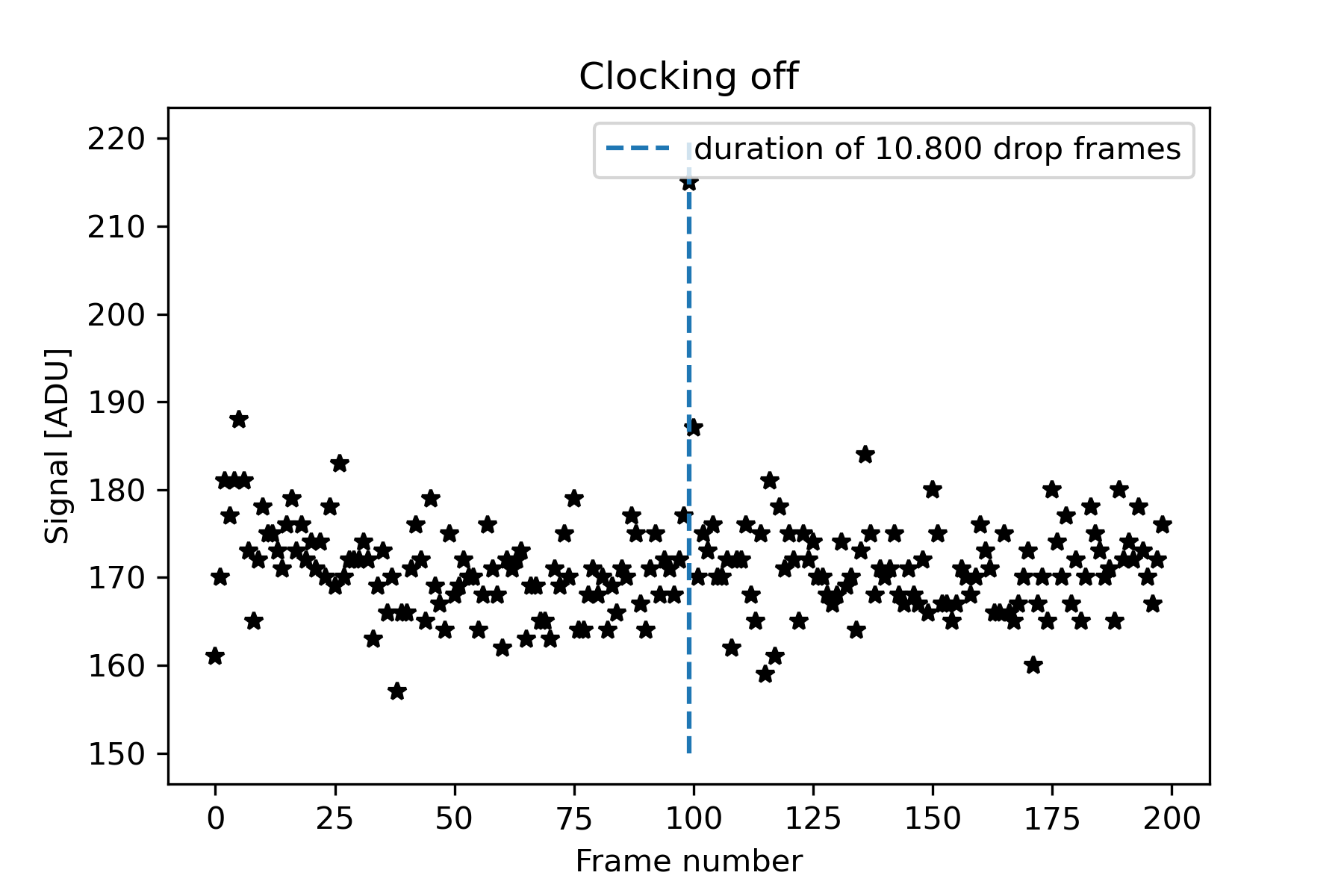}
  \end{subfigure}
  \caption{Fowler-sampling measurement of the glow and DC signals generated during an 11.000 frame test (here represented for one pixel). \textit{Left:} Glow test where drops frames are taken. The high number of drop frames of this exposure allows the array to integrate a consistent amount of glow between both series of reads and accurately measure it as a function of number of frames. \textit{Righ:} Same test without the drop frames. Only dark-current is here integrated which, over the timescale of the test ($\sim$2h) is too low to be noticed on such a plot. The signal fluctuations are read noise.}
  \label{fig:glow test}
\end{figure*}

To measure both glow and intrinsic dark-current separately, we take two Fowler-sampled data sets of identical exposure time:

\begin{itemize}
    \item In the first dataset illustrated for one pixel on Fig.\ref{fig:glow test}, we acquire drop frames during the exposure, which means the array is clocked and periodically probed during the acquisition as it would be if read frames were taken like in an up-the-ramp (UTR) sampling fashion. This dataset is thus affected by dark-current and glow (generated at every read or drop frame).
    \item  The second dataset is taken without clocking (no drop frames). The detector is still powered but not solicited during the exposure, meaning that only the dark current is contributing.
\end{itemize}

\begin{figure}
\begin{center}
\begin{tabular}{c}
\includegraphics[height=5.5cm]{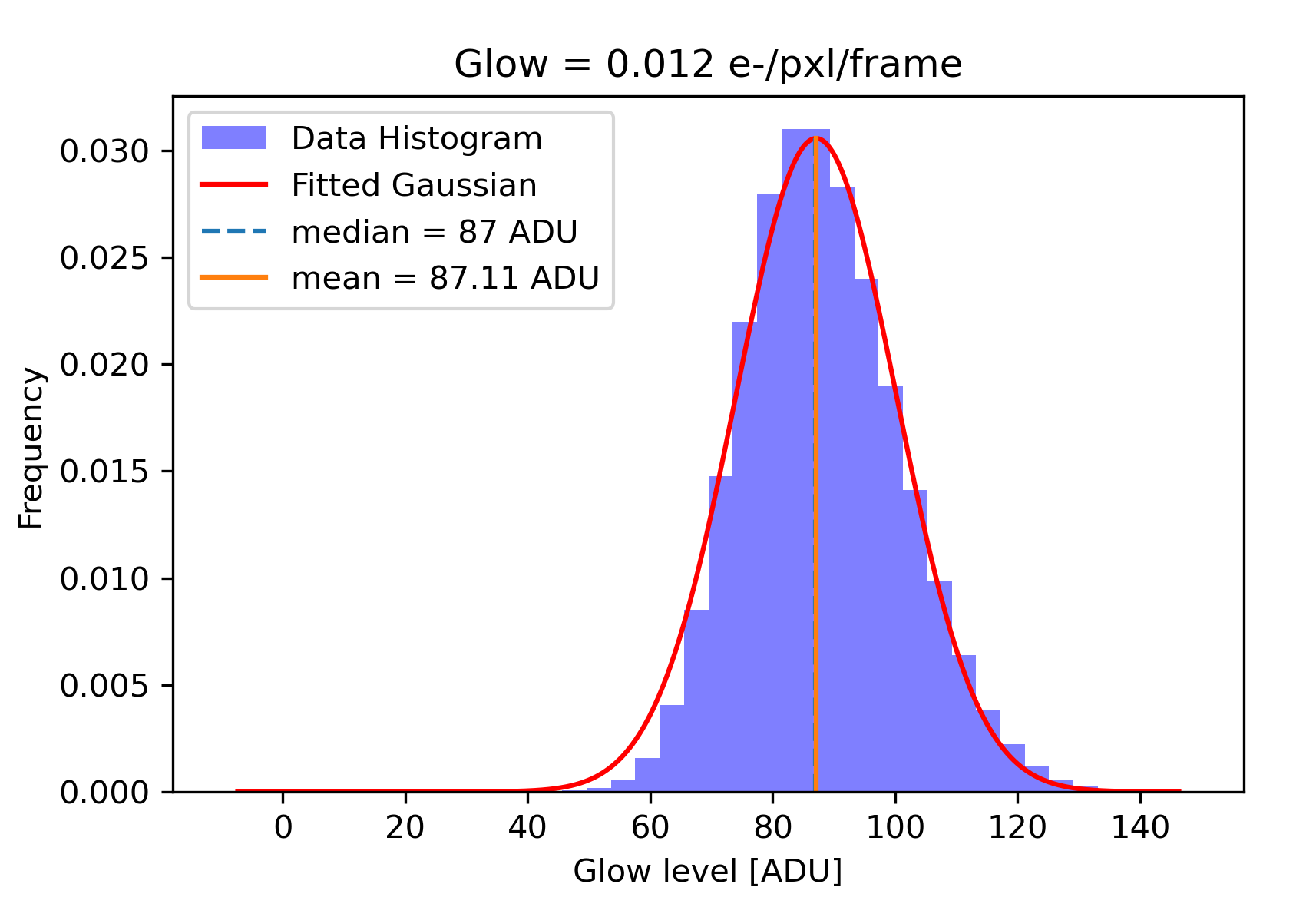}
\end{tabular}
\end{center}
\caption 
{ \label{fig:histo}
Glow histogram of the entire array.}
\end{figure}

\begin{figure*}[!tbp]
\centering
    \begin{subfigure}[b]{0.49\textwidth}
    \centering
    \includegraphics[width=\textwidth]{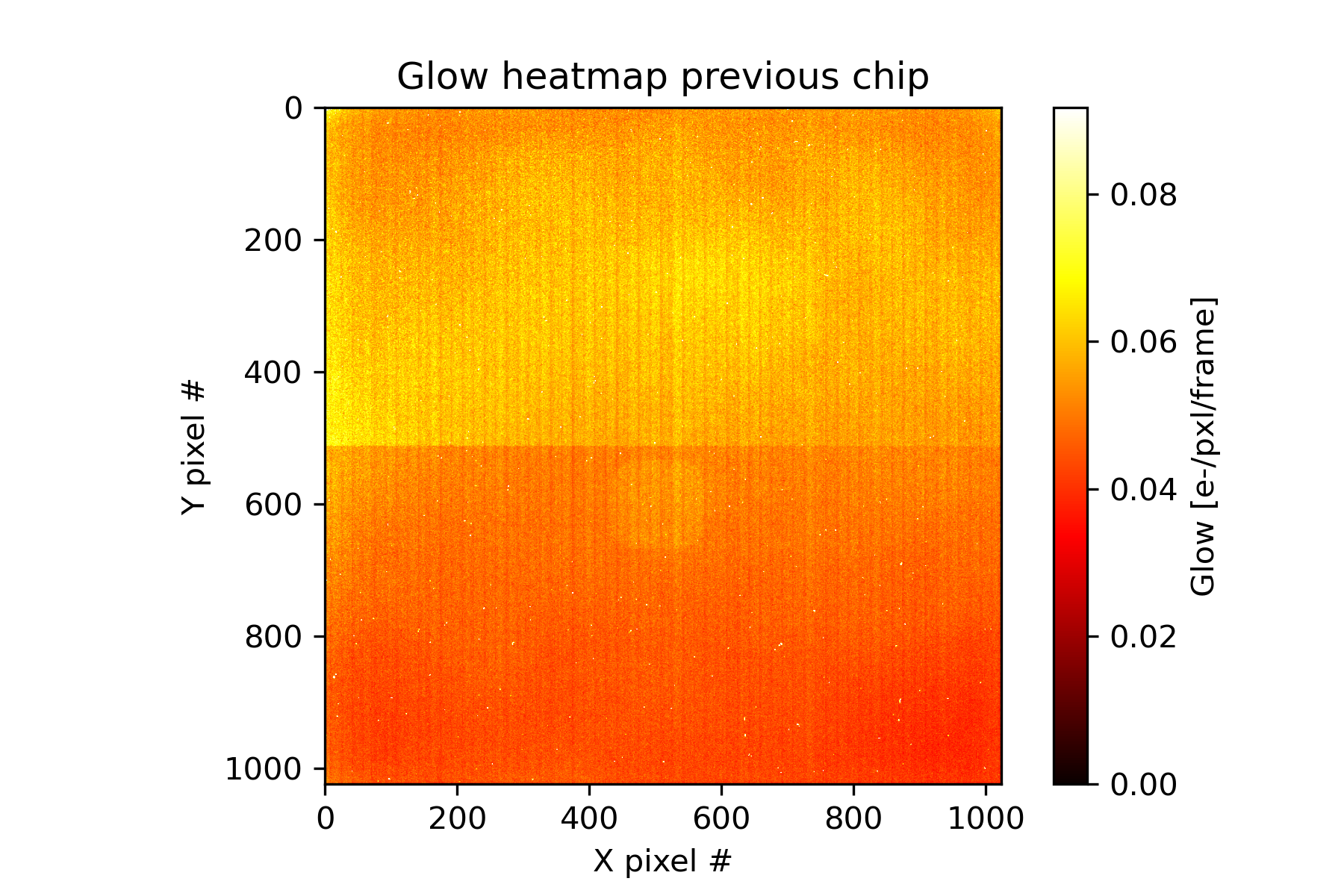}
  \end{subfigure}
  \hfill
  \begin{subfigure}[b]{0.49\textwidth}
    \centering
    \includegraphics[width=\textwidth]{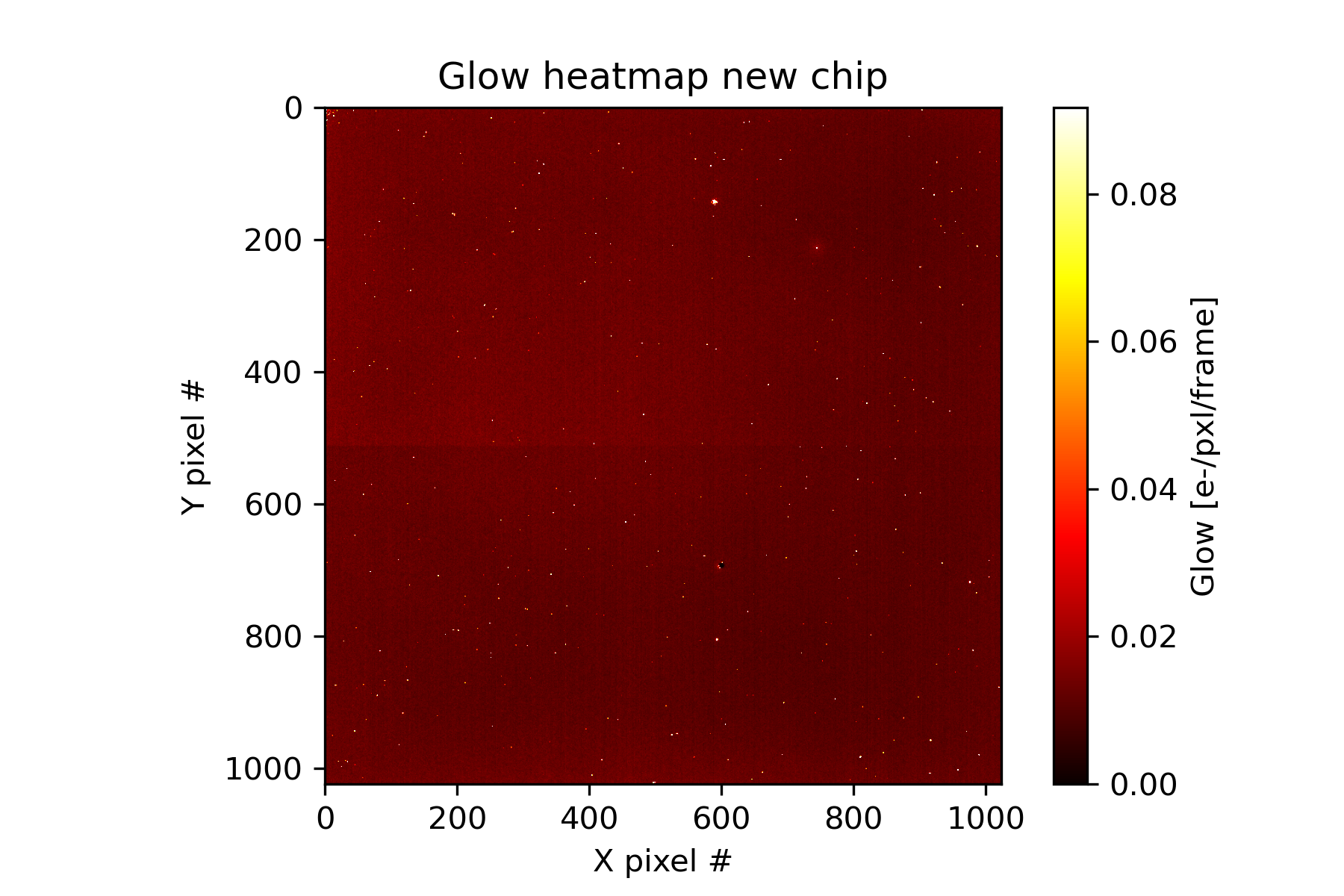}
  \end{subfigure}
  \caption{Glow heatmaps with same intensity scales. \textit{Left:} heatmap of the previous array characterized in C22; \textit{Right:} array studied in this paper. The up/down asymmetry is due to the architecture of the detector. To limit the amount of glow received by the previous chip, we were using a small mask that appears on the heatmap. We suspect the glow in this area to be reflected off the edges of the mask and more generally the detector chamber, thus increasing the measured signal within the mask.}
  \label{fig:heatmaps}
\end{figure*}

Both tests are composed of 11.000 frames ($2\times100$ read frames and 10.800 drop frames in between). The observed glow in the first test can be observed on Fig.\ref{fig:glow test} where see a clear increase of signal between both series of read frames. In the case of the second test, no increase is observed and the intrinsic dark current over the whole detector is essentially 0. To measure the actual glow level, we subtract the medians of both series of 100 read frames to average the read noise. Averaging over the whole array, we measure an overall glow of 0.012e-/pixel/frame (\textit{cf.} Fig.\ref{fig:histo}), which is $\sim 7$ times smaller than the value from the previous sensor reported in C22. We can have a better look at the per pixel glow by plotting a heatmap of the array. Fig.\ref{fig:heatmaps} shows how the glow has been corrected in the new device compared to previous one.

\begin{figure}
\begin{center}
\begin{tabular}{c}
\includegraphics[height=5.5cm]{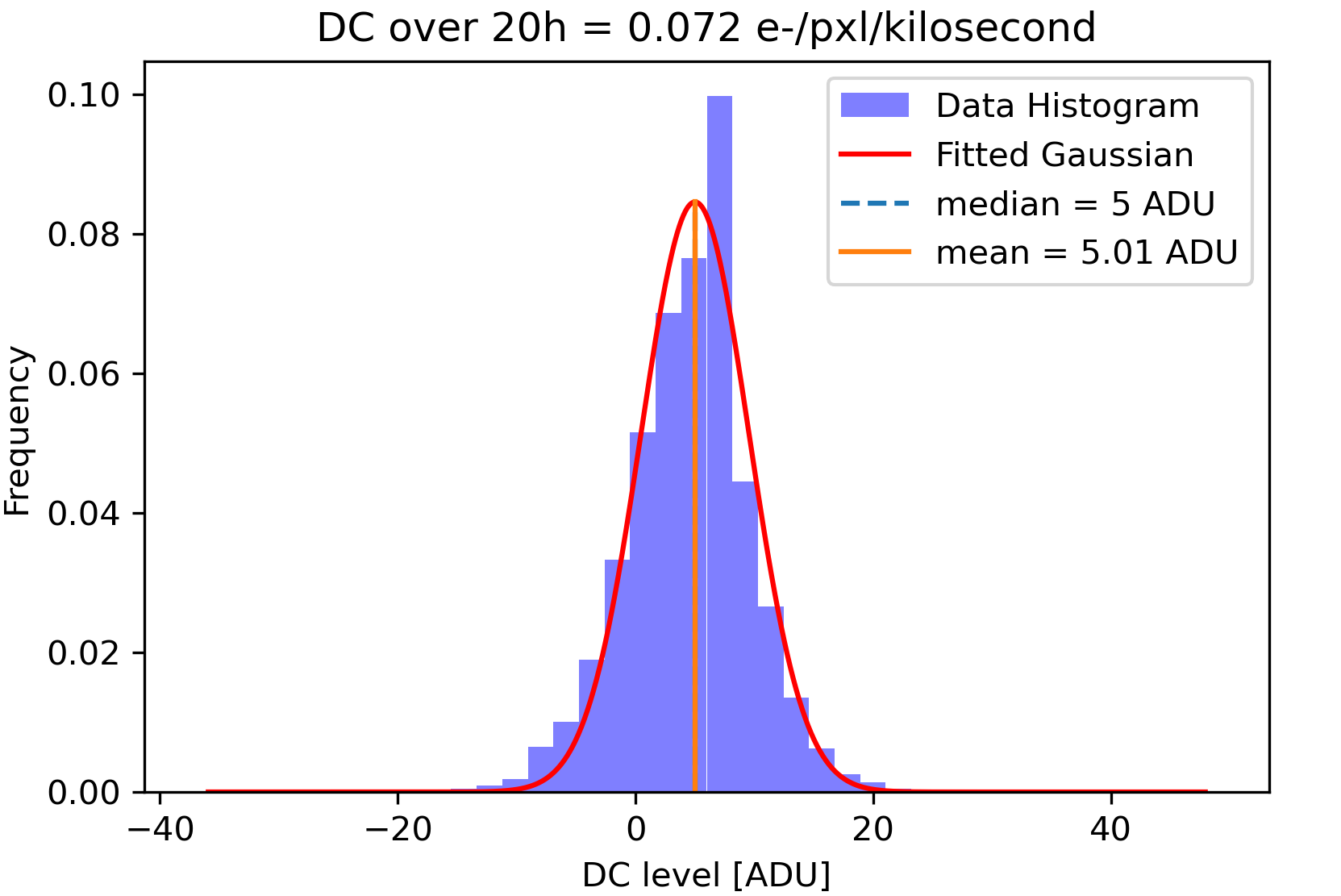}
\end{tabular}
\end{center}
\caption 
{ \label{fig:DC histo}
Dark-current histogram of the entire array. Over a 20 hours test, the mean DC signal is only 5 ADU $\sim$ 7.7 e-.}
\end{figure}

Now that the glow has been determined, we can measure the dark current with the second test, accounting for the small amount of glow being integrated during the read frames. Two hours (the duration of the 11.000 frames test) was actually too short to obtain a significant dark-current measurement as seen on Fig.\ref{fig:glow test} so a second test of 20 hours was done (Fig.\ref{fig:DC histo}). We obtain a dark current of $\sim 0.07$ e-/pixel/kilosecond which is essentially consistent with a null dark current considering usual exposure times. A brief summary of our results in comparison with values from C22 can be found in Table \ref{tab:summary}.

\begin{table*}[ht]
\caption{Summary table of our results relative to  our previous work C22.} 
\begin{center}

\begin{tabular}{|c || c | c |} 
 \hline
  & C22 & This work  \\ [0.5ex] 
 \hline\hline
 Gain [e-/ADU] & 1.51 $\pm$ 0.02 & 1.54 $\pm$ 0.01\\ 
 \hline
 Glow [e-/pixel/frame] & 0.08 & 0.012 $\pm$ 0.001\\ 
 \hline
 Effective DC [e-/pixel/kilosecond] & $\sim$ 0.1 & 0.07 $\pm$ 0.03  \\
 \hline
\end{tabular}
\end{center}
\label{tab:summary}
\end{table*}

\section{Individual photon detection tests}\label{single}

\subsection{Photon jumps}

Now that glow and dark-current have been evaluated, we operate at high amplification gain to study the sensitivity in the very low flux regime. In particular, we show that LmAPDs are capable of detecting single photons. Detectors with this ability have been know for decades, but only recently have they been implemented in imaging devices such as CMOS and CCDs \cite{2012A&A...537A..50H, 2015NatCo...6.5913M, 2019LSA.....8...87B}. Until now, such technologies have been limited to optical wavelengths and shorter, no mature imaging device has been shown to perform such demands in the infrared while having low noise. At high bias voltage, photoelectrons are highly amplified by the avalanche process and the effective read noise becomes smaller than the signal of a single photoelectron. Glow and dark-current do not fully experience the amplification process, enabling to distinguish individual photon events.

\begin{figure}
\begin{center}
\begin{tabular}{c}
\includegraphics[height=5.7cm]{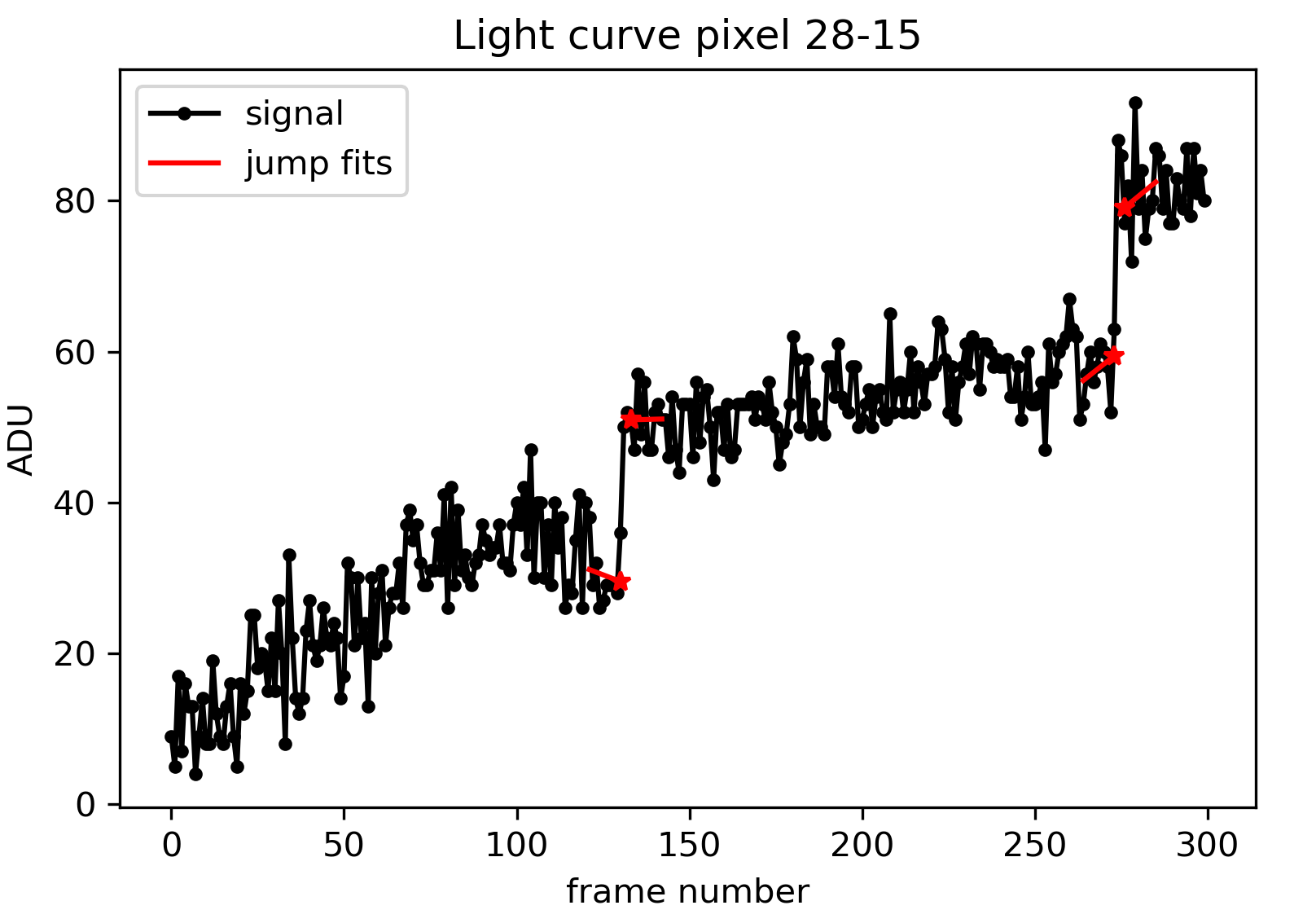}
\end{tabular}
\end{center}
\caption 
{ \label{fig:single photon}
Light curve of a pixel at ultra-low flux. The overall trend is dark-current and glow, the fluctuations are read noise. The jumps correspond to individual photon detections.}
\end{figure}

To operate in this regime, we set the bias voltage at 12V where the effective read noise is below 1 e-/pixel/frame. To minimize the dark current, we run the detector up-the-ramp at high frequency ($\sim$0.7 frame/s) in a  300 read frames test. The light source has been attenuated such that in average, a pixel receives only a few photons during the duration of the test. Fig.\ref{fig:single photon} shows the example of a well-functioning pixel where the signal of the photoelectrons is notably above the read noise level. The slow general increase is due to dark-current and glow.

Most ($\sim$80\%) pixels visually display photon detection capabilities. However, in order to get the most out of this data, it is useful to work with larger datasets which we can statistically analyze. To do so, we extended the test to 6000 frames and we implemented an Python routine that detects photoelectron signals in the light curves and measures their amplitude. We use a changepoint detection algorithm to detect the photoelectron jumps, and measure the jump amplitude by fitting two affine functions left and right of the jump as seen on Fig.\ref{fig:single photon}.

We believe this approach can be a new way to measure many parameters of the detector such as glow, dark-current and read-noise by studying the "quiet" sections of the light-curves, as well as measuring the gain and excess noise factor via the jumps properties.

\subsection{Relation with glow}

Since the pixels under the mask do not receive any illumination, it is insightful to look at their behavior in this test to see if any jumps are observed, in which case they would be attributed to glow photons.

We find very few jump detections under the mask, many of them with smaller amplitudes than the photon jumps in the illuminated region. In particular, in a given time series, we detect fewer jumps than the expected number of glow events based on the number of frames of the test by one order of magnitude. This observation suggests that glow photons are not fully amplified, they could be captured within the amplification region. If glow photons are only partially amplified, that makes them distinguishable from signal photons, which is encouraging.

\section{Summary and future work}
\label{sec:summary}

In conclusion, this work demonstrates the capabilities of LmAPDs to operate in an ultra-low background imaging environment. The reduction of the glow on the latest version of the detectors leads to an intrinsic dark current $<<1$ e-/pixel/kilosecond at 50K and 4V of bias voltage, limited by a per-frame ROIC glow signal of $\sim 0.01$ e-/pixel/frame. This enables photon counting performance at high bias-voltage, which ultimately is the end goal of these devices.

As many detectors, LmAPDs manifest persistence which will be characterized in the near future. We also plan on doing radiation testing to see how the sensors would behave in space, in the presence of cosmic rays. Finally, it will be interesting to investigate how the individual photon detection tests can be used as a new mean of characterizing detector properties.

\acknowledgments 
 
The authors would like to thank Ian Baker from Leonardo for his helpful insights and advice. Development of the detectors and their characterization at UH is sponsored by NASA SAT award \#18-SAT18-0028.

\bibliography{report} 
\bibliographystyle{spiebib} 

\end{document}